\documentclass[final,5p,times,nonatbib]{elsarticle}
\usepackage[T1]{fontenc}
\usepackage[utf8]{inputenc}
\usepackage[english]{babel}
\usepackage{csquotes}
\usepackage{amsmath}
\usepackage{booktabs,array,tabularx,longtable}
\usepackage{graphicx}
\usepackage{xcolor}
\usepackage{tikz}
\usetikzlibrary{arrows.meta,positioning,calc,fit}
\makeatletter
\@ifclassloaded{elsarticle}{\let\c@author\relax}{}
\makeatother
\usepackage[backend=biber,style=apa,sorting=nyt,doi=true,url=true,isbn=false]{biblatex}
\AtBeginDocument{}
\usepackage[hidelinks]{hyperref}
\usepackage{microtype}
\usepackage{balance}
\definecolor{metablue}{HTML}{205B78}
\definecolor{metateal}{HTML}{277F80}
\definecolor{metaorange}{HTML}{A45E20}
\newcolumntype{Y}{>{\raggedright\arraybackslash}X}
\newcolumntype{P}[1]{>{\raggedright\arraybackslash}p{#1}}

\AtEveryBibitem{\clearfield{issn}}

\journal{Computers \& Education: X Reality}
\date{24 October 2026}
\begin{document}
\begin{frontmatter}
\title{Designing collaborative virtual environments for software engineering education}
\author[1]{Filipe Fernandes}
\ead{filipe.fernandes@ifsudestemg.edu.br}
\author[2]{Cláudia Werner}
\ead{werner@cos.ufrj.br}
\affiliation[1]{organization={Department of Computer Science -- Federal Institute of Southeast of Minas Gerais (IF Sudeste MG)},
  city={Manhuaçu},
  state={MG},
  country={Brazil}}
\affiliation[2]{organization={Systems Engineering and Computer Science -- Federal University of Rio de Janeiro (UFRJ)},
  city={Rio de Janeiro},
  state={RJ},
  country={Brazil}}
\begin{abstract}
Collaborative virtual environments offer opportunities to make software models available for shared inspection, but educational design and usability need to be examined separately from claims of learning effectiveness. This article presents MetaSEE, an approach connecting requirements guidance, a Web XR feature model, a reusable UML component, and a browser-based multiuser platform. We synthesize documented formative evaluations of the framework (16 participations), feature model (15), UML component (9), and platform (14). These counts represent evaluation participations; the same 12 developers took part in the first two evaluations. The principal platform study involved six teachers working individually and eight students working in four pairs through desktop or notebook computers, keyboard, and mouse. The documentary synthesis traces educational intentions to implemented features and reported difficulties. Participants identified potential value in shared modeling and simulation, alongside navigation, interaction, visualization, and distraction concerns. A numerical audit reconstructs a student System Usability Scale mean of 67.50 from published item frequencies, conflicting with the reported 79.38; an incomplete teacher item distribution prevents recovery of that group's overall mean. The contribution is a traceable account of educational design and formative limitations, with practical implications for onboarding, model readability, collaboration, and evaluation reporting. The evidence supports bounded feasibility and perceived usefulness, without establishing learning gains, sustained engagement, or adoption.
\end{abstract}
\end{frontmatter}
\section{Introduction}
\label{sec:introduction}

Software engineering education asks students to coordinate representations, decisions, and people. A class diagram, for example, represents a domain through classes, attributes, operations, and relationships, but its educational value also depends on how students explain and revise those choices. A modeling exercise therefore involves both the artifact and the discussion around it. Conventional teaching can provide worked examples and diagram editors; arranging opportunities for learners to examine a shared model and negotiate its meaning remains an instructional design problem. Research on software engineers' knowledge gaps also cautions against equating curriculum coverage with preparation for practice \parencite{Garousi2019}. This motivates educational activities that make disciplinary reasoning visible, without assuming that any particular technology will improve it.

Three-dimensional virtual environments offer one possible setting for such activities. They can place learners and manipulable representations in a shared space, combining alternative viewpoints with communication around an artifact. The learning affordances identified by \textcite{DalgarnoLee2010} include spatial knowledge representation, experiential and contextualized activity, motivation, and collaborative learning. These are possibilities arising through suitable tasks and interactions. They do not establish that a three-dimensional display or an avatar is educationally effective in itself. \textcite{Fowler2015} develops this point through a design-for-learning perspective: pedagogical activity needs explicit attention alongside the technical characteristics of the environment.

This distinction matters particularly for software models. Increasing the available display space may support inspection, yet it can also introduce navigation, selection, and text-entry demands. The software visualization literature treats interaction and evaluation as central issues alongside visual representation \parencite{Teyseyre2009}. More directly, collaborative UML modeling in virtual reality has already been investigated through VmodlR, including a comparison with a conventional web modeling tool \parencite{Yigitbas2023}. Thus, the relevant research question is no longer whether UML can be represented in a virtual environment. It is how a particular educational design connects a modeling activity, the conditions of participation, the available tools, and the evidence obtained when users try them.

The current literature provides complementary answers at different levels. SENEM links educational requirements to a prototype and iterative usability evaluation \parencite{Pentangelo2024}. VoRtex emphasizes support for constructing collaborative educational experiences \parencite{Jovanovic2022}. A recent scoping review in this journal connects the design, technological arrangements, and outcomes of virtual reality applications in computer science education \parencite{Janani2026}. These contributions support examining both the design rationale and the quality of the available evidence. A favorable reaction to a prototype, successful manipulation of an object, and improvement in conceptual understanding answer different questions; reporting them as equivalent would obscure what educators can reasonably infer.

MetaSEE addresses the design and development of virtual environments for software engineering education through a requirements framework, a feature-based approach, a reusable UML component called VirtualUML, and an integrated prototype platform. Earlier reviews characterized software engineering applications and immersive learning frameworks \parencite{FernandesWerner2022,FernandesFrameworks2023}. The present article builds on those foundations by explaining the educational rationale of the artifacts and synthesizing the formative evaluations documented in the thesis by \textcite{FernandesThesis2023}. Its additional contribution is the connection between design decisions, activities, and observed barriers. The reviews themselves, architectural aspirations, and features that remained unimplemented are distinguished from the evaluated platform.

Four evaluations provide the source evidence. E1 examined the framework, E2 the feature model, E3 VirtualUML, and E4 the integrated platform. E4 is the principal evaluation because it involved the intended educational users: six teachers working individually and eight students working in four pairs. Participants accessed the platform remotely through desktop or notebook computers with keyboard and mouse. This is the interaction modality to which the findings apply. Possibilities involving immersive headsets form part of the broader design space and future work. E1--E3 contribute complementary perspectives from developers and academics rather than independent replications of E4.

Three research questions organize the present documentary synthesis of evaluations originally planned through Goal Question Metric:
\begin{description}
 \item[RQ1.] How does MetaSEE connect educational and technological requirements to the design of virtual environments for software engineering education?
 \item[RQ2.] What design contributions and limitations are indicated by evaluations of the framework, feature model, and VirtualUML?
 \item[RQ3.] What usability perceptions and barriers emerge from teachers' and students' use of the platform?
\end{description}
These questions clarify the scope of the current account; they are not presented as prospectively registered hypotheses or as the wording originally used in the evaluations.

The article makes three related contributions. First, it presents an explicit mapping from educational objectives and activities to implemented resources and unresolved requirements. Second, it offers a traceable formative synthesis that retains critical feedback and separates evidence about each artifact. Third, it audits the reported usability aggregates and explains which numerical conclusions can be recovered from the available documentation. The resulting account supports decisions about feasibility, perceived usefulness, interaction, and further development. It does not demonstrate learning gains, sustained engagement, or adoption. In particular, E4 has no comparison condition or learning assessment, the teacher and student tasks differ, and the individual questionnaire responses are unavailable for the present synthesis. These boundaries determine both the interpretation of the findings and the priorities for subsequent educational evaluation.

\section{Related work and pedagogical grounding}
\label{sec:related-work}

\subsection{From three-dimensional representations to learning activities}

A virtual environment offers an arrangement for learning rather than a pedagogy by itself. The distinction is consequential for software engineering, where the visible objects represent abstractions that learners must interpret. \textcite{DalgarnoLee2010} connect representational fidelity and learner--computer interaction to potential learning benefits. For MetaSEE, three of their affordances are especially relevant: inspecting spatial representations, performing contextualized activities, and collaborating around shared objects. Motivation is a further possibility to investigate, not a consequence established by the presence of avatars. This affordance perspective is used here as an analytical lens for the present synthesis; it was not an educational theory experimentally tested in the original evaluations.

The design-for-learning argument of \textcite{Fowler2015} helps translate that lens into practical questions. What should a learner do, what representation or interaction makes that activity possible, and what observation would indicate its educational value? For a UML task, rotating or traversing a representation can enable inspection. Creating and revising classes can provide practice in expressing a domain. Discussing a shared model can give participants opportunities to justify alternatives. These are our proposed instructional connections. Evidence that an interface permits the actions is different from evidence that the actions improved understanding, and both differ from evidence that participants found the experience appealing.

The challenge is not restricted to virtual environments. Software visualization research examines how representations, interaction techniques, and evaluation methods support software-related tasks \parencite{Teyseyre2009}. A representation that is technically renderable may still be difficult to read or manipulate. For this reason, our educational mapping includes barriers alongside intended affordances. Navigation, object selection, and visibility are relevant because they can interrupt a modeling discussion. The analysis asks whether the interface makes the disciplinary activity practicable, while leaving claims about understanding and transfer to studies that measure those outcomes.

\subsection{Modeling and process-oriented software engineering environments}

Earlier educational applications illustrate different ways of making software engineering activity visible. VisAr3D provides a three-dimensional representation of UML models \parencite{Rodrigues2016}. It is a relevant visualization antecedent, whereas MetaSEE's account concerns the combination of a reusable modeling resource, a shared environment, and a broader approach to application development. This is a distinction in the scope of the designs; it does not demonstrate that one representation is more effective than another.

VmodlR is a particularly close comparator. \textcite{Yigitbas2023} evaluated collaborative class-diagram construction with 24 participants in 12 pairs, using both their virtual reality environment and Lucidchart. Participants used headsets and controllers for the virtual reality condition. Most pairs took longer with that condition, despite favorable aspects of the experience. This evidence prevents a simple inference from spatial interaction to improved efficiency. It also identifies a methodological difference: MetaSEE's E4 used desktop interaction and provides formative observations without a comparison condition. We therefore do not rank the systems by usability scores or generalize across these modalities.

Process-oriented applications address another educational purpose. Scrum VR places learners in a simulated software team, with interactions concerning Scrum roles and practices \parencite{Mayor2021}. Its initial teacher and student evaluations were used to inform development. A subsequent ScrumVR study included 78 software engineering students and pre--post assessment \parencite{LopezFernandez2023}. That distinction matters: a study with knowledge assessment can address questions that a usability questionnaire alone cannot answer. MetaSEE's contribution concerns the design and usability of a modeling environment, and should be assessed at that level.

\subsection{Educational platforms and design frameworks}

SENEM is a close comparison for the passage from requirements to an evaluated educational platform. \textcite{Pentangelo2024} combine a literature survey and educator/learner interviews, derive nine requirements, implement a prototype, and apply Rapid Iterative Testing and Evaluation to learnability, efficiency, and satisfaction. Its title refers to software engineering as an enabling approach to development; it should not be read as evidence that the platform specifically teaches software engineering. Our comparison instead concerns requirements, user involvement, and evaluation scope.

VoRtex provides a multiuser platform and tools for creating gamified collaborative educational experiences \parencite{Jovanovic2022}. Its reported evaluation includes comparison of platform capabilities through Manninen's matrix. Such an architectural comparison is useful for identifying supported activities, but it is a different form of evidence from a learning assessment. \textcite{Roy2023} also contribute a review-based framework for the educational metaverse. These works inform the design space; their presence does not imply that a particular MetaSEE component, interoperability mechanism, or analytics facility was implemented.

The newer MITE framework connects requirements investigation, infrastructure, and learning content through explicit pedagogical models \parencite{BezzinaZammit2026}. Its proof of concept is accompanied by limitations concerning independent field testing and educational evaluation. This reinforces the value of distinguishing guidance for design from demonstrated outcomes. MetaSEE contributes documented feedback across several related artifacts, while its present pedagogical mapping remains an interpretation of that record. Table~\ref{tab:related-comparison} compares selected close designs at this level rather than treating all educational metaverses as interchangeable.

\begin{table*}[t]
\caption{Selected related designs: capabilities and reported evidence are distinguished. This is a focused comparison, not an exhaustive inventory.}
\label{tab:related-comparison}
\centering\small
\begin{tabularx}{\textwidth}{P{0.14\textwidth}YYY}
\toprule
Design & Educational or design focus & Reported resource or activity & Evidence relevant to comparison \\
\midrule
VmodlR \parencite{Yigitbas2023} & Collaborative UML modeling & Shared class-diagram creation in headset VR & Paired within-participant comparison with web modeling; time, operating errors, and perceptions \\
Scrum VR / ScrumVR \parencite{Mayor2021,LopezFernandez2023} & Scrum roles and practices & Scenario-based serious game & Formative user evaluations; later study includes pre--post knowledge assessment \\
SENEM \parencite{Pentangelo2024} & Requirements for educational environments & Implemented educational platform & Requirements elicitation and iterative evaluation of interaction qualities \\
VoRtex \parencite{Jovanovic2022} & Gamified collaborative learning & Multiuser platform and content-building support & Comparison of prototype capabilities with other virtual world platforms \\
MITE \parencite{BezzinaZammit2026} & Pedagogically grounded engineering environments & Design methodology and proof of concept & Initial application; larger and independent educational evaluations remain necessary \\
MetaSEE (present synthesis) & SE environment design and UML activity & Framework, feature approach, VirtualUML, and desktop platform & Documentary synthesis of E1--E4; perceptions and barriers, without demonstrated learning gains \\
\bottomrule
\end{tabularx}
\end{table*}

\subsection{Positioning the present evidence}

The recent computing-education review by \textcite{Janani2026} includes 30 studies representing 29 virtual reality applications and identifies variation in topics, interaction arrangements, theoretical grounding, and assessment. It also highlights the scarcity of longitudinal evidence. That diversity makes modality, task, and outcome definitions essential to comparison. We use the review to update the educational context rather than to infer MetaSEE's effectiveness from findings obtained with other applications.

The earlier reviews associated with MetaSEE provide foundations for requirements and framework development \parencite{FernandesWerner2022,FernandesFrameworks2023}. The contribution added here is the traceable relation between those concerns, implemented resources, and formative findings. Evaluation constructs are kept separate: the System Usability Scale concerns perceived usability \parencite{Brooke1996,Lewis2018}; perceived usefulness and ease of use belong to the acceptance literature \parencite{Davis1989}. E3's acceptance-inspired items retain that limited status. E4 is not retroactively described as a Technology Acceptance Model evaluation. This positioning permits useful design conclusions while preserving the distinction between what users could do, what they reported, and what they learned.

\section{Research design}
\label{sec:research-design}

\subsection{Design trajectory and scope of this synthesis}

The research followed a design science trajectory in which problems identified in educational practice informed the construction and formative evaluation of artifacts \parencite{Hevner2004,Peffers2007}. The artifacts comprised an educational design framework, a feature model for web-based extended reality applications, a reusable UML component, and a platform integrating shared virtual worlds with software modelling facilities. Their purposes differed: the framework prompted design decisions, the feature model represented application variability, the component supported diagram construction, and the platform exposed these facilities to teachers and students. Evaluating one artifact therefore did not validate the others or establish the effectiveness of the resulting educational activity.

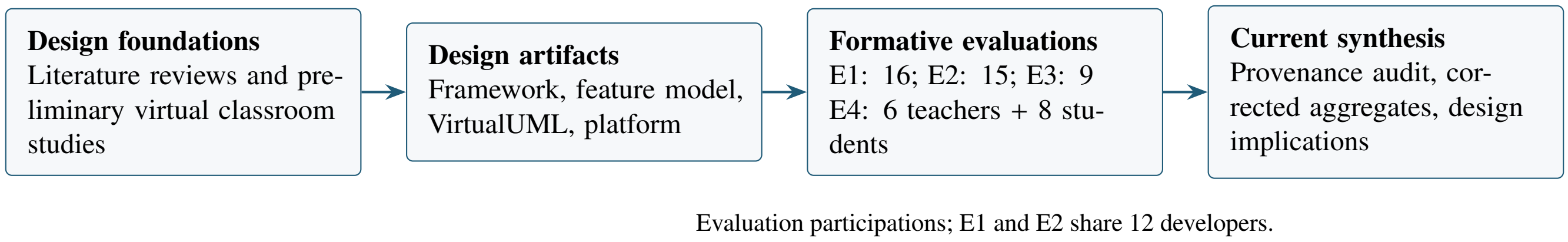
\begin{figure*}[t]
\centering
\begin{tikzpicture}[font=\small,box/.style={draw=metablue,rounded corners=2pt,fill=metablue!4,text width=3.5cm,minimum height=1.45cm,align=left,inner sep=7pt},arr/.style={-{Stealth},thick,metablue}]
\node[box] (a) {\textbf{Design foundations}\\Literature reviews and preliminary virtual classroom studies};
\node[box,right=5mm of a] (b) {\textbf{Design artifacts}\\Framework, feature model, VirtualUML, platform};
\node[box,right=5mm of b] (c) {\textbf{Formative evaluations}\\E1: 16; E2: 15; E3: 9\\E4: 6 teachers + 8 students};
\node[box,right=5mm of c] (d) {\textbf{Current synthesis}\\Provenance audit, corrected aggregates, design implications};
\draw[arr] (a)--(b);\draw[arr](b)--(c);\draw[arr](c)--(d);
\node[below=3mm of c,align=center,font=\footnotesize] {Evaluation participations; E1 and E2 share 12 developers.};
\end{tikzpicture}
\caption{Research trajectory and the boundary of the present contribution. The synthesis revisits documented evaluations without collecting new participant data. The preliminary classroom study is an antecedent, separate from E1--E4.}
\label{fig:trajectory}
\end{figure*}

The present article synthesizes the evaluations documented in the doctoral research record \parencite{FernandesThesis2023}, with no new participant recruitment or data collection. The original studies were planned through goals, questions and metrics (GQM), relating an evaluation goal to questions and response measures \parencite{Basili1994}. The three research questions in this article organize the current documentary synthesis; they were not prospectively registered experimental hypotheses for the 2023 sessions. We use \emph{formative evaluation} to describe their evidential role, although the original reports called them experiments. None used random allocation, a comparison condition or a design that estimates the causal effect of MetaSEE on learning.

Earlier work with a virtual classroom prototype supplied design experience and procedural lessons. That antecedent included a four-participant pre/post assessment of software requirements knowledge. Three participants obtained the same result before and after using the prototype, and one obtained a lower post-test result. It supplies no evidence of learning gain. It is kept separate from E1--E4 because it involved a different prototype, task and purpose. Its reported usability summary is likewise not substituted for a MetaSEE platform result.

\subsection{Evaluation corpus and participants}

The evidence corpus comprises the thesis's evaluation narratives, questionnaire instruments, response-frequency figures, summary tables and recorded participant extracts. The predecessor manuscript provides a consistency check on the platform description and reporting, rather than an independent replication. E1 reviewed the framework with four academics and twelve developers (16 participations); E2 reviewed the feature model with three academics and the same twelve developers (15 participations); E3 evaluated VirtualUML with four academics and five developers (nine participations). E4 involved six teachers and eight students (14 participants). These counts describe participation in each evaluation. They cannot be summed as 54 distinct people, and the archive does not support a verified total of unique participants across all evaluations.

\begin{table*}[t]
\centering\small
\caption{Evaluation corpus and evidential purpose. Counts refer to participation in each evaluation, excluding its pilot.}
\label{tab:evaluations}
\begin{tabularx}{\textwidth}{@{}P{1.8cm}P{3.1cm}YY@{}}
\toprule
Evaluation & Participants & Object and procedure & Evidential scope \\
\midrule
E1 & 4 academics; 12 developers & Framework scenario/review and importance ratings & Requirements guidance and clarity; no measured requirement-quality gain \\
E2 & 3 academics; same 12 developers & Feature-model inspection and feature descriptions & Perceived correspondence and coverage; no generator benchmark \\
E3 & 4 academics; 5 developers & Component activity / documentation review & Usefulness and ease-of-use perceptions under unequal exposure \\
E4 & 6 teachers individually; 8 students in 4 pairs & Desktop class-modeling task, SUS, open questions & Situated perceived usability and barriers; no learning assessment \\
\bottomrule
\end{tabularx}
\end{table*}

Academics and developers were recruited through professional networks and research-oriented communication channels. Their E1--E3 instruments differed, including shorter documentation or statement reviews for developers. E4 used a convenience sample of teachers and volunteer students from one Brazilian educational institution. Consequently, the evidence represents situated judgments and use experiences rather than a representative estimate for software engineering educators or learners. Pilot sessions informed refinements and are described separately from the main evaluation counts.

\subsection{E4 procedure and instruments}

E4 focused on perceived usability during a UML class-modelling activity. A two-person pilot in April 2023 preceded the main sessions in the same month. The pilot used individual remote sessions, verbalized observations and questionnaire feedback. Its revisions are reported in Section~\ref{sec:e4-pilot}. The main evaluation comprised six individual teacher sessions through Google Meet and four student-pair sessions through a private Discord server. Pairs were arranged according to availability. External meeting applications supported communication and observation; their facilities should not be read as evaluated features of MetaSEE.

Participants used desktop or notebook computers to access the browser-based platform, navigating with a keyboard and controlling the view with a mouse. The instruments recommended Chrome or Firefox and closing other applications. They requested operating-system, memory, download-speed and browser information, but the available report does not establish a controlled hardware configuration. The study did not evaluate headsets, tracked controllers or immersive display conditions.

Four task stages structured each session: open the platform, explore freely for up to ten minutes, enter a virtual world, and construct an analysis class diagram. Students entered the same world in pairs and interpreted a written educational-domain scenario. Teachers worked individually from a supplied diagram. The task representations and collaboration conditions therefore differed. Table~\ref{tab:e4-tasks} specifies these differences so that usability responses are understood in relation to the activity actually performed.

Instruments were administered in Portuguese. The protocol contained a consent form, a characterization form, profile-specific execution instructions and a post-task evaluation form. The consent instrument covered adult participation, confidentiality and the right to withdraw. The evaluation combined ten System Usability Scale (SUS) items with open questions about the modelling experience, desired functionality, and advantages and disadvantages of virtual worlds. The characterization questions about classroom engagement concerned participants' prior educational contexts; they were not measures of engagement caused by MetaSEE. E4 did not administer the usefulness and ease-of-use items used in E3 as a separate acceptance instrument.

\subsection{Documentary synthesis and numerical audit}

The synthesis first mapped each finding to its artifact, evaluation, participant profile and source location. It then distinguished observed use, questionnaire judgments, expectations and proposed extensions. For qualitative evidence, we retained the thesis's categories and participant extracts, including unfavorable reports. We did not reconstruct a full interview corpus, recode recordings, claim thematic saturation or calculate agreement between coders. Category counts are reported as documented mentions; the archive does not establish mutually exclusive categories or a one-to-one mapping between positive and negative summaries.

The numerical audit checked denominators and transcribed the published SUS charts. SUS scoring subtracts one from odd-numbered responses, subtracts even-numbered responses from five, and multiplies the sum by 2.5 \parencite{Brooke1996,Lewis2018}. Because this transformation is linear, complete item-level frequencies can recover a group mean without recovering individual scores. They cannot recover the covariance needed for respondent-level score variability or reliability. All student item totals were complete; the teacher chart contained only five responses on its first item for six participants. We report the student mean derivable from those aggregates, identify its disagreement with the historical text, and leave the overall teacher mean unrecoverable. No missing value, confidence interval, reliability estimate or between-group test is manufactured. The supplement records the transcription and calculation for inspection.

\section{MetaSEE educational design}
\label{sec:design}

\subsection{From educational requirements to a bounded prototype}

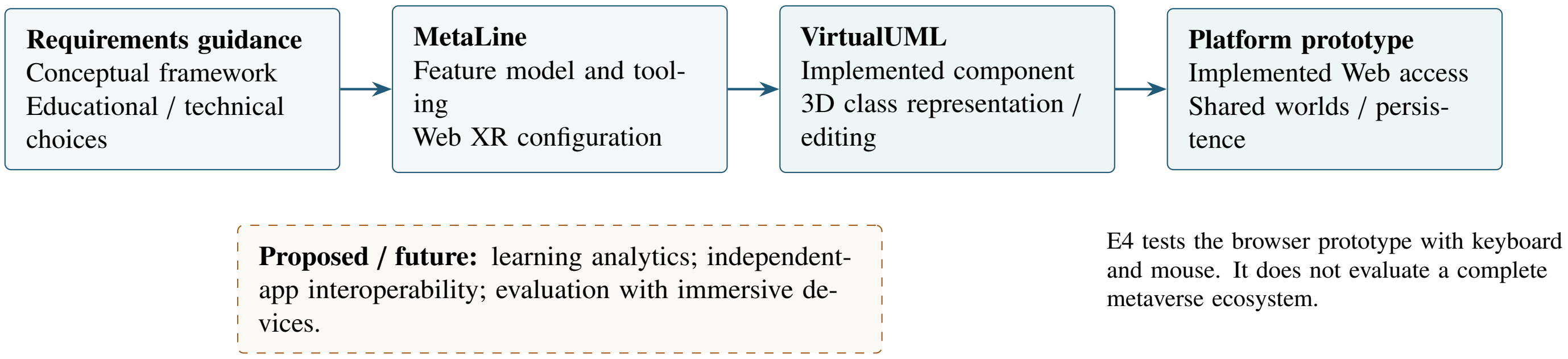
\begin{figure*}[t]
\centering
\begin{tikzpicture}[font=\small,box/.style={draw=metablue,rounded corners=2pt,text width=3.4cm,minimum height=1.4cm,align=left,inner sep=7pt},arr/.style={-{Stealth},thick,metablue}]
\node[box,fill=metablue!5] (a) {\textbf{Requirements guidance}\\Conceptual framework\\Educational / technical choices};
\node[box,fill=metablue!5,right=6mm of a] (b) {\textbf{MetaLine}\\Feature model and tooling\\Web XR configuration};
\node[box,fill=metateal!8,right=6mm of b] (c) {\textbf{VirtualUML}\\Implemented component\\3D class representation / editing};
\node[box,fill=metateal!8,right=6mm of c] (d) {\textbf{Platform prototype}\\Implemented Web access\\Shared worlds / persistence};
\draw[arr](a)--(b);\draw[arr](b)--(c);\draw[arr](c)--(d);
\node[below=6mm of b,draw=metaorange,dashed,rounded corners=2pt,fill=metaorange!4,align=left,text width=7cm,inner sep=7pt] (f) {\textbf{Proposed / future:} learning analytics; independent-app interoperability; evaluation with immersive devices.};
\node[below=6mm of d,align=left,text width=5.3cm,font=\footnotesize] {E4 tests the browser prototype with keyboard and mouse. It does not evaluate a complete metaverse ecosystem.};
\end{tikzpicture}
\caption{Compact view of MetaSEE with explicit implementation status. Arrows indicate the relationship of design artifacts; they do not imply that end-to-end automated generation or independent-application interoperability was empirically validated.}
\label{fig:approach}
\end{figure*}

MetaSEE connects three kinds of design work: specifying an educational activity, configuring the technology through which learners encounter it, and implementing software-engineering objects that can be manipulated in a shared world. These levels address different users. An educator needs to define the modeling problem and organise participation; a developer needs to select interaction mechanisms and implement representations; a learner needs to understand the task while navigating and editing. The design therefore includes both development artifacts and a browser platform. Figure~\ref{fig:approach} distinguishes these artifacts from the broader architecture proposed around them.

The original requirements emerged from earlier reviews of immersive-learning frameworks and virtual environments for software-engineering education \parencite{FernandesFrameworks2023,FernandesWerner2022}, together with a preliminary virtual-classroom prototype. They included coverage of software-engineering topics, alternative visualization and interaction, support for application development, analysis of learning data, and interoperability. Their implementation was uneven. The evaluated platform supports a specific class-modeling activity, while analytics and independent-application interoperability remain proposals. Presenting that distinction is essential because a design requirement describes something sought, whereas a working feature and an educational result require separate evidence.

The educational target instantiated here is constructing and discussing a simple analysis class diagram. Learners identify candidate classes and links in a problem description and externalise them as objects in a shared space. The activity offers an opportunity to negotiate a representation with a partner. The evaluation records participants' experiences with that opportunity; it does not establish whether negotiation occurred productively, whether diagrams were correct, or whether knowledge improved.

\subsection{Making pedagogical and technological choices explicit}

The MetaSEE Framework comprises 19 guiding questions in ten areas: immersion, user experience, software visualization, development tools, application features, learning indicators, learning outcomes, pedagogical approaches, student profile, and context. The framework asks designers to identify educational content and intended competencies alongside practical choices such as devices, interaction techniques, supported users, and execution conditions. Its purpose is to prompt decisions that might otherwise be implicit. It does not prescribe one pedagogy or guarantee that a completed checklist yields a suitable activity.

For the UML instantiation, the software representation is a class diagram, the participation arrangement is shared modeling, and the access mode is a conventional computer browser. The use of a 3D scene makes position and viewpoint part of the interaction. It also adds navigation and camera-control demands that a conventional diagram editor may avoid. Consequently, an educational rationale must specify what learners should do together with the representation and examine whether the interface permits that action without excessive effort.

Table~\ref{tab:design-alignment} reconstructs this relationship for the present synthesis. It is an analytical interpretation of documented requirements, features, and evaluations, rather than a pedagogical theory tested in the original studies. E1's criticism that the framework mixed educational goals with implementation choices motivates keeping the columns separate. The resulting mapping makes the evidential boundary visible: a feature can instantiate an intended activity even when the intended learning outcome remains unmeasured.

\begin{table*}[t]
\centering
\caption{Educational intentions, activities, implemented resources, and available evidence. The mapping is an interpretation developed for this article.}
\label{tab:design-alignment}
\small
\begin{tabularx}{\textwidth}{@{}P{0.18\textwidth}YYY@{}}
\toprule
Educational intention & Activity & Documented resource & Evidence and boundary \\
\midrule
Externalise a software model & Create classes and connect them to represent a scenario & VirtualUML class and association entities; graphical creation/editing interface & E3 component perceptions and E4 use; no scored diagram-quality outcome \\
Support discussion around an artifact & Student pairs work in the same virtual world & Networked multiuser scene with a shared model & E4 pair sessions and recorded perceptions; no interaction-quality or learning measure \\
Make educational design decisions explicit & Specify topic, intended outcomes, learners, and context & Framework questions alongside technology-related prompts & E1 formative feedback, including genericity and mixed-concern criticism \\
Reduce repeated configuration work & Select application features and generate a starting skeleton & MetaLine feature model and editor & E2 examines model correspondence; productivity and generator correctness were not measured \\
Enable access using available computers & Enter a world through a browser and use familiar input devices & Portal, shareable world address, keyboard/mouse interaction & Desktop E4 only; no evidence about headset use or equal access across devices \\
\bottomrule
\end{tabularx}
\end{table*}

\subsection{Configuring reusable application elements}

MetaLine represents variability in a family of Web XR applications. Its model describes platforms, interfaces, XR modalities, devices, browsers, avatars, virtual worlds, and optional interaction events. The associated editor translates selected features into an A-Frame code skeleton. This separates recurring scene and input configuration from the educational content of a particular application. A developer can start with selected components and then implement the activity-specific behaviour.

The distinction between a feature model and a verified implementation matters here. A model can include wearable or mobile options without demonstrating that every possible combination works or supports learning. E2 elicited judgments about the model and its description, including uncertainty about completeness and requests for clearer device-specific capabilities. It did not compare development time or test generated applications across the modeled device space. MetaLine is therefore reported as a documented development artifact with formative model feedback.

\subsection{VirtualUML and the platform boundary}

VirtualUML provides A-Frame entities for class representations and associations. Its reusable component accepts properties such as a class name and position, while associations refer to the identifiers of their endpoints. A platform extension adds a graphical menu and editing panel, so platform users need not write these scene definitions themselves. Keeping the rendering component separate from platform integration allows the representation to be reused without embedding platform-specific persistence in every component.

The implementation supports a limited subset of UML modeling. The component evaluation identifies additional relationships, stereotypes, attributes, methods, and other UML diagram types as requested extensions. The extension screenshot nevertheless contains fields labelled Attributes and Operations, while its depicted class object displays a name. The record does not reconcile those interface fields with the component-level requests or establish the extent of working notation support. Accordingly, the screenshots and task description substantiate the creation and arrangement of classes and links; the presence of an input field is insufficient evidence of a complete UML authoring environment. This restriction narrows the educational activity to simple structural representations and limits transfer to richer modeling assignments.

The browser portal lists worlds and provides search, account access, world creation, and entry links. Worlds have shareable addresses. Within a world, network functionality supports multiple users, and the platform's API stores class information, including its name, ownership, position, rotation, scale, and colour. These mechanisms connect the portal, shared scene, and persistence within the prototype. They do not demonstrate data exchange among independently developed applications. Likewise, the architecture's learning-analytics pipeline describes intended collection, integration, analysis, and feedback; that module was not implemented.

\subsection{A concrete UML activity}

\begin{figure*}[t]
\centering
\begin{minipage}[t]{.485\textwidth}
\textbf{1. Edit the model's class content}\par\smallskip
\begin{tikzpicture}
\node[inner sep=0] (img) {\includegraphics[width=\linewidth]{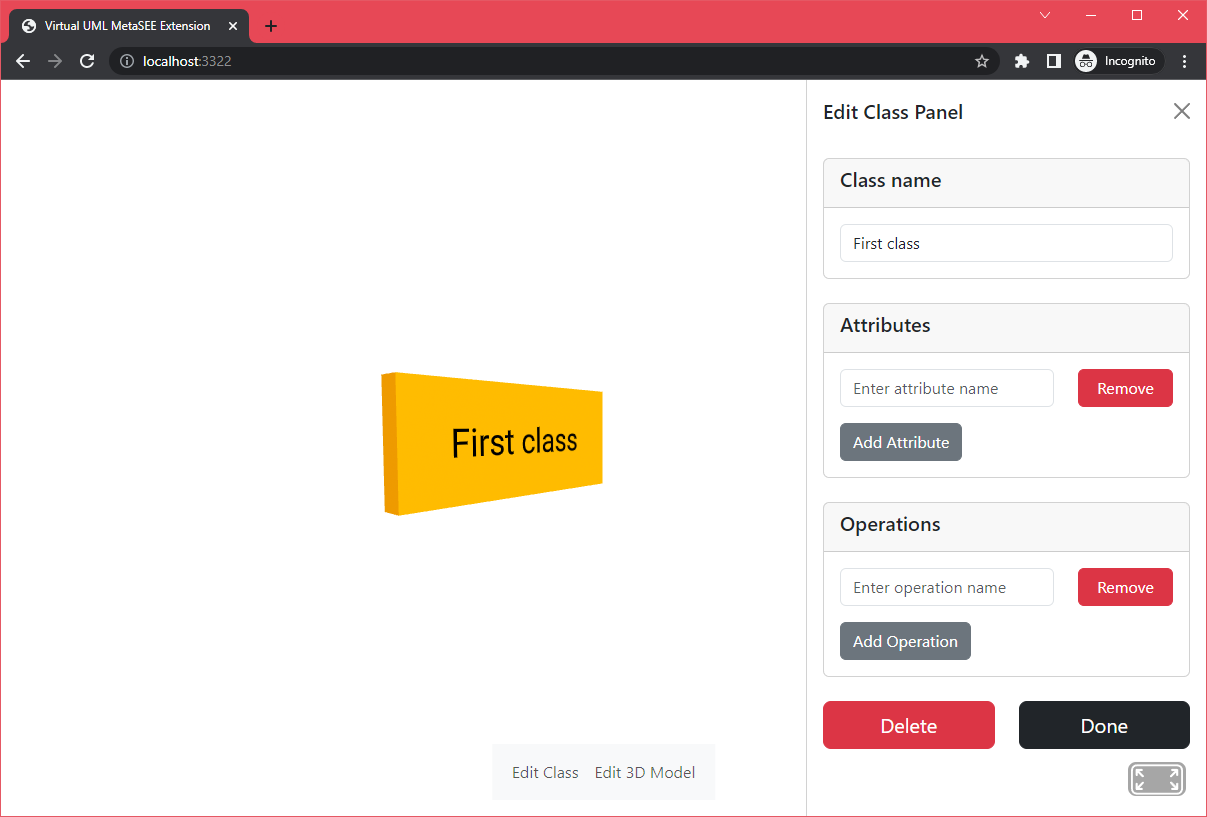}};
\begin{scope}[shift={(img.south west)},x={($(img.south east)-(img.south west)$)},y={($(img.north west)-(img.south west)$)}]
\node[fill=metablue,text=white,circle,font=\bfseries\small,inner sep=3pt] at (.70,.78) {A};
\node[fill=metablue,text=white,circle,font=\bfseries\small,inner sep=3pt] at (.70,.43) {B};
\end{scope}
\end{tikzpicture}
\footnotesize A: class name. B: attributes/operations input fields; complete rendering support is not established. Existing prototype capture.
\end{minipage}\hfill
\begin{minipage}[t]{.485\textwidth}
\textbf{2. Inspect the classes in the shared world}\par\smallskip
\begin{tikzpicture}
\node[inner sep=0] (img) {\includegraphics[width=\linewidth]{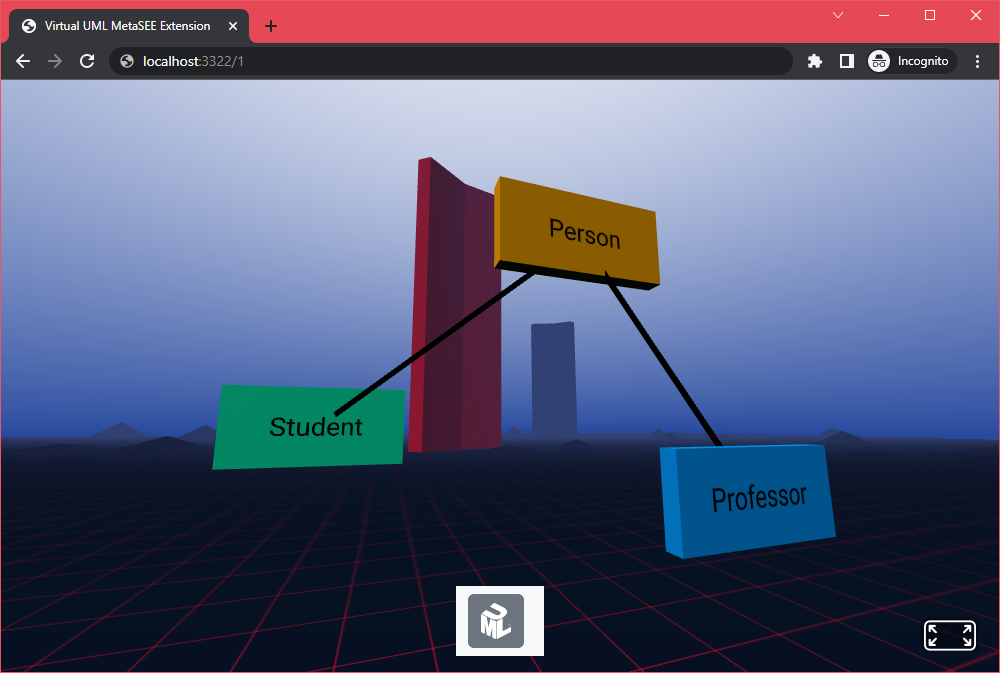}};
\begin{scope}[shift={(img.south west)},x={($(img.south east)-(img.south west)$)},y={($(img.north west)-(img.south west)$)}]
\node[fill=metablue,text=white,circle,font=\bfseries\small,inner sep=3pt] at (.55,.66) {C};
\node[fill=metablue,text=white,circle,font=\bfseries\small,inner sep=3pt] at (.49,.10) {D};
\end{scope}
\end{tikzpicture}
\footnotesize C: visible Person--Student--Professor example. D: modeling menu. Students performed modeling in pairs; teachers modeled individually.
\end{minipage}
\caption{Illustrative task sequence assembled from documented prototype screenshots, with explanatory annotations added. Editing class content and inspecting the 3D model are shown separately to clarify the user workflow. The screenshot example is not the complete E4 scenario; the exact prescribed task is specified in the methods and supplement. No new interface functionality is depicted.}
\label{fig:uml}
\end{figure*}

The documented user journey begins with entering the browser platform and freely exploring it for up to ten minutes. Participants then enter a virtual world, create class objects through the VirtualUML menu, edit their information or appearance, and connect objects using associations. Figure~\ref{fig:uml} illustrates these operations with existing prototype captures. The annotations explain the interaction shown and do not add unavailable functions.

In E4, teachers individually reproduced a supplied six-class diagram concerning property and financing. Its reference image contains the Portuguese labels for property, financing, land, residential property, apartment, and house, joined by simple lines. Students instead worked in pairs from a prose scenario involving teachers, subjects, courses, classes, and students. Their task required discussing which classes and links would express the scenario, while the teachers' task supplied a visual target. These are related but distinct modeling demands. Neither arrangement provides a basis for ranking the groups' modeling ability or comparing their usability perceptions as if task conditions were equivalent.

The interface also required users to control an avatar and viewpoint while working with the model. Pilot feedback led to placing newly created classes closer to the avatar, clarifying a menu label, indicating the class being edited, and reducing the clicks needed to create associations. These are concrete examples of formative evaluation changing access to the educational activity. The main evaluation still identified interaction and visualization barriers, making navigation, selection feedback, and diagram readability continuing design concerns.

General scene and access mechanisms could in principle be reused for other subject areas, but such transfer would require different disciplinary objects and tasks. The documented evaluations concern software-engineering design and UML activities; they establish no cross-domain educational validity.

The evaluated resource is thus a browser-based collaborative modeling prototype. Its contribution is a traceable account of how an educational task is connected to requirements, reusable representations, and interface decisions, together with evidence about difficulties encountered. Headsets, advanced analytics, broader UML coverage, and cross-application exchange require implementation and evaluation before they can extend that contribution.

\section{Formative evaluation}
\label{sec:formative-evaluation}

\subsection{Complementary evidence from E1--E3}

E1--E3 supplied judgments about the artifacts supporting the platform. Their value lies in identifying design considerations before interpreting E4 as a situated use experience. In E1, the four academics were divided on the framework's organization and clarity: two agreed and two disagreed on each aspect. Three agreed that it covered basic development considerations. Reported weaknesses included generic guidance and insufficient separation of pedagogical decisions from implementation choices. Developer ratings concerned the importance of the nineteen guidelines, rather than observed success in developing an educational application. Favorable relevance judgments therefore coexist with the need for more concrete examples and clearer sequencing.

E2 reviewed the feature model with three academics and the twelve developers also involved in E1. All three academics endorsed its potential for extension, while two were neutral about completeness. Developer agreement with feature descriptions indicates recognition of the proposed scope; it does not demonstrate successful product derivation or interoperability. In E3, academics performed a creation-oriented task involving VirtualUML and its documentation, whereas developers reviewed documentation. Most reported favorable usefulness and ease-of-use judgments, but the recorded limitations included UML complexity, incomplete notation coverage and insufficient coordinate explanations. These findings support refining the modelling facilities and their instructional support. They do not measure student learning, and the acceptance-oriented questions from E3 cannot be attributed to E4.

The evaluation formats explain these boundaries. E1 academics first considered an application-design scenario and then revisited it with the framework; developers rated the importance of guidelines through a shorter form. These procedures elicited views about decisions worth considering. They did not compare the quality or cost of independently implemented educational applications. One academic recommended starting with functional requirements and integration, then addressing non-functional concerns such as security, performance and budget. This suggestion helps operationalize the criticism of generic guidance: designers need an order for making decisions and examples connecting them to their particular educational task.

E2 used textual descriptions of feature-model choices to avoid assuming familiarity with feature-modelling notation. The resulting agreement should consequently be interpreted as judgments about the described choices, rather than demonstrated competence in configuring the model. Academics also requested clearer diagrams and more explicit mappings between devices, platforms and available capabilities. E3 raised a related distinction between a reusable component and an application ready for a teacher. Academics worked with code and an online editor, and unfamiliarity with the editor was already identified in the pilot. Developers' documentation-based judgments did not reproduce those practical conditions. Across the three evaluations, the recurring issue was the support required to translate an available technical facility into an understandable educational activity.

\subsection{E4 pilot and main task context}
\label{sec:e4-pilot}

The pilot made navigation experience an explicit design concern. One participant identified keyboard movement and mouse camera controls intuitively and related this familiarity to gaming experience. The other, without comparable game experience, encountered navigation difficulties. The study consequently added game-experience questions to the characterization instrument. This two-person observation identifies a plausible source of interaction variation; it does not establish a statistical association or explain all subsequent differences between teachers and students.

The pilot also produced four changes reported as implemented before E4: placing newly created classes closer to the avatar, replacing the misleading appearance-editing menu wording, indicating which class was being edited, and reducing the number of clicks needed to create associations. These revisions connect participant feedback to concrete interaction decisions. They should be distinguished from improvements requested after the main sessions, for which implementation was not documented.

\begin{table*}[t]
\centering
\caption{E4 task conditions. The common platform did not make the two groups' activities equivalent.}
\label{tab:e4-tasks}
\small
\begin{tabularx}{\textwidth}{@{}P{0.14\textwidth}YY@{}}
\toprule
Stage & Teachers (six individual sessions) & Students (four pairs; eight participants) \\
\midrule
Access & Open the browser-based platform on a desktop or notebook & Open the same platform on a desktop or notebook \\
Familiarization & Explore freely for up to ten minutes & Explore freely for up to ten minutes \\
World entry & Enter a virtual world individually & Enter the same virtual world as the partner \\
Modelling task & Construct an analysis class diagram from a supplied visual model & Jointly interpret a written domain scenario and construct an analysis class diagram \\
Remote session & Google Meet & Private Discord server, selected for simultaneous screen sharing \\
Post-task evidence & Individual SUS and open responses & Individual SUS and open responses following paired activity \\
\bottomrule
\end{tabularx}
\end{table*}

The student prompt described a teacher teaching several subjects, each subject belonging to one teacher, subjects belonging to courses, and classes composed of students. This scenario required identifying domain classes and expressing relationships in an analysis diagram. It supplied material for collaborative interpretation rather than a measured test of UML knowledge. The teachers instead received a six-class property-and-financing diagram with links to reproduce. Neither a common rubric for diagram correctness nor a controlled task-completion measure is recoverable from the reported results. We therefore use the task descriptions to explain the interaction context, without converting participation into evidence of accurate models, improved conceptual understanding or superior performance by either group.

\subsection{Audited SUS evidence}
\label{sec:e4-sus}

The thesis reported mean SUS values of 81.25 for teachers and 79.38 for students. The associated frequency figures do not support carrying those summaries forward as verified results. Student frequencies contain eight responses for each of the ten items. Applying standard item-direction scoring to these frequencies yields 216 adjusted response points over the group and a derived mean of $2.5\times216/8=67.50$. This is the value recoverable from the documented aggregates, conditional on their accuracy. It does not establish that the unavailable individual response file was correctly represented by the chart. The difference from the historical 79.38 remains an unresolved reporting discrepancy.

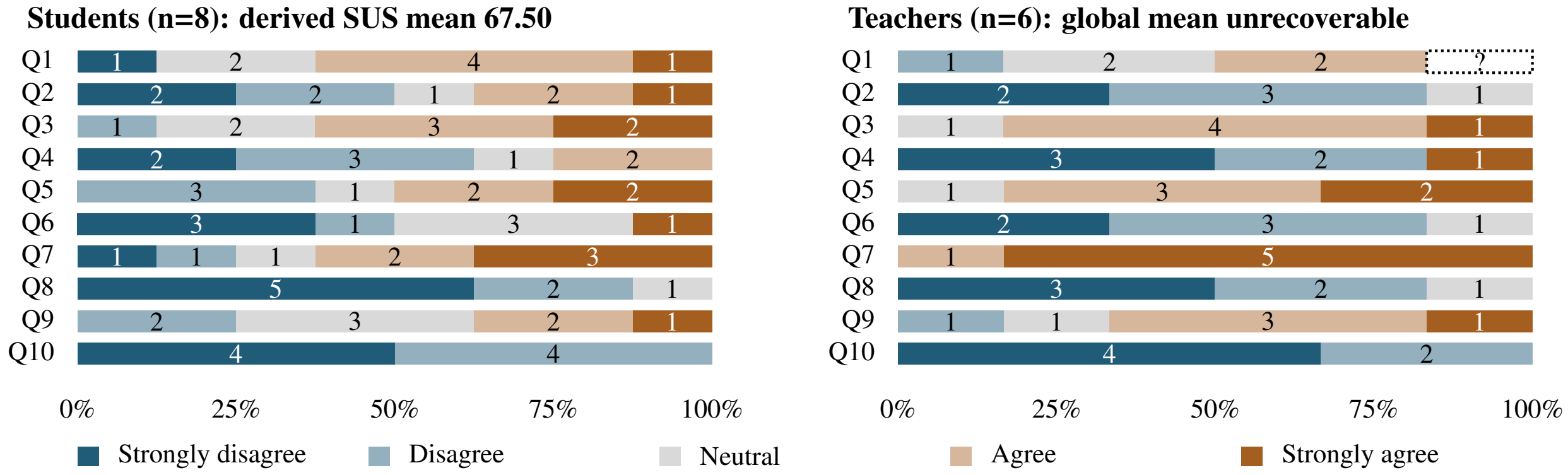
\begin{figure*}[t]
\centering
\begin{tikzpicture}[x=1cm,y=1cm,font=\footnotesize]
\begin{scope}[shift={(0,0)}]
\node[anchor=west,font=\small\bfseries] at (-.65,.55) {Students (n=8): derived SUS mean 67.50};
\node[anchor=east] at (-.12,0.115) {Q1};
\fill[metablue] (0.0000,0.0000) rectangle (0.8313,0.2300);
\node[text=white] at (0.4156,0.1150) {1};
\fill[black!15] (0.8313,0.0000) rectangle (2.4938,0.2300);
\node[text=black] at (1.6625,0.1150) {2};
\fill[metaorange!45] (2.4938,0.0000) rectangle (5.8188,0.2300);
\node[text=black] at (4.1562,0.1150) {4};
\fill[metaorange] (5.8188,0.0000) rectangle (6.6500,0.2300);
\node[text=white] at (6.2344,0.1150) {1};
\node[anchor=east] at (-.12,-0.22500000000000003) {Q2};
\fill[metablue] (0.0000,-0.3400) rectangle (1.6625,-0.1100);
\node[text=white] at (0.8313,-0.2250) {2};
\fill[metablue!48] (1.6625,-0.3400) rectangle (3.3250,-0.1100);
\node[text=black] at (2.4938,-0.2250) {2};
\fill[black!15] (3.3250,-0.3400) rectangle (4.1562,-0.1100);
\node[text=black] at (3.7406,-0.2250) {1};
\fill[metaorange!45] (4.1562,-0.3400) rectangle (5.8187,-0.1100);
\node[text=black] at (4.9875,-0.2250) {2};
\fill[metaorange] (5.8187,-0.3400) rectangle (6.6500,-0.1100);
\node[text=white] at (6.2344,-0.2250) {1};
\node[anchor=east] at (-.12,-0.5650000000000001) {Q3};
\fill[metablue!48] (0.0000,-0.6800) rectangle (0.8313,-0.4500);
\node[text=black] at (0.4156,-0.5650) {1};
\fill[black!15] (0.8313,-0.6800) rectangle (2.4938,-0.4500);
\node[text=black] at (1.6625,-0.5650) {2};
\fill[metaorange!45] (2.4938,-0.6800) rectangle (4.9875,-0.4500);
\node[text=black] at (3.7406,-0.5650) {3};
\fill[metaorange] (4.9875,-0.6800) rectangle (6.6500,-0.4500);
\node[text=white] at (5.8188,-0.5650) {2};
\node[anchor=east] at (-.12,-0.905) {Q4};
\fill[metablue] (0.0000,-1.0200) rectangle (1.6625,-0.7900);
\node[text=white] at (0.8313,-0.9050) {2};
\fill[metablue!48] (1.6625,-1.0200) rectangle (4.1562,-0.7900);
\node[text=black] at (2.9094,-0.9050) {3};
\fill[black!15] (4.1562,-1.0200) rectangle (4.9875,-0.7900);
\node[text=black] at (4.5719,-0.9050) {1};
\fill[metaorange!45] (4.9875,-1.0200) rectangle (6.6500,-0.7900);
\node[text=black] at (5.8187,-0.9050) {2};
\node[anchor=east] at (-.12,-1.245) {Q5};
\fill[metablue!48] (0.0000,-1.3600) rectangle (2.4938,-1.1300);
\node[text=black] at (1.2469,-1.2450) {3};
\fill[black!15] (2.4938,-1.3600) rectangle (3.3250,-1.1300);
\node[text=black] at (2.9094,-1.2450) {1};
\fill[metaorange!45] (3.3250,-1.3600) rectangle (4.9875,-1.1300);
\node[text=black] at (4.1562,-1.2450) {2};
\fill[metaorange] (4.9875,-1.3600) rectangle (6.6500,-1.1300);
\node[text=white] at (5.8188,-1.2450) {2};
\node[anchor=east] at (-.12,-1.5850000000000002) {Q6};
\fill[metablue] (0.0000,-1.7000) rectangle (2.4938,-1.4700);
\node[text=white] at (1.2469,-1.5850) {3};
\fill[metablue!48] (2.4938,-1.7000) rectangle (3.3250,-1.4700);
\node[text=black] at (2.9094,-1.5850) {1};
\fill[black!15] (3.3250,-1.7000) rectangle (5.8188,-1.4700);
\node[text=black] at (4.5719,-1.5850) {3};
\fill[metaorange] (5.8188,-1.7000) rectangle (6.6500,-1.4700);
\node[text=white] at (6.2344,-1.5850) {1};
\node[anchor=east] at (-.12,-1.925) {Q7};
\fill[metablue] (0.0000,-2.0400) rectangle (0.8313,-1.8100);
\node[text=white] at (0.4156,-1.9250) {1};
\fill[metablue!48] (0.8313,-2.0400) rectangle (1.6625,-1.8100);
\node[text=black] at (1.2469,-1.9250) {1};
\fill[black!15] (1.6625,-2.0400) rectangle (2.4938,-1.8100);
\node[text=black] at (2.0781,-1.9250) {1};
\fill[metaorange!45] (2.4938,-2.0400) rectangle (4.1562,-1.8100);
\node[text=black] at (3.3250,-1.9250) {2};
\fill[metaorange] (4.1562,-2.0400) rectangle (6.6500,-1.8100);
\node[text=white] at (5.4031,-1.9250) {3};
\node[anchor=east] at (-.12,-2.265) {Q8};
\fill[metablue] (0.0000,-2.3800) rectangle (4.1562,-2.1500);
\node[text=white] at (2.0781,-2.2650) {5};
\fill[metablue!48] (4.1562,-2.3800) rectangle (5.8187,-2.1500);
\node[text=black] at (4.9875,-2.2650) {2};
\fill[black!15] (5.8187,-2.3800) rectangle (6.6500,-2.1500);
\node[text=black] at (6.2344,-2.2650) {1};
\node[anchor=east] at (-.12,-2.605) {Q9};
\fill[metablue!48] (0.0000,-2.7200) rectangle (1.6625,-2.4900);
\node[text=black] at (0.8313,-2.6050) {2};
\fill[black!15] (1.6625,-2.7200) rectangle (4.1562,-2.4900);
\node[text=black] at (2.9094,-2.6050) {3};
\fill[metaorange!45] (4.1562,-2.7200) rectangle (5.8187,-2.4900);
\node[text=black] at (4.9875,-2.6050) {2};
\fill[metaorange] (5.8187,-2.7200) rectangle (6.6500,-2.4900);
\node[text=white] at (6.2344,-2.6050) {1};
\node[anchor=east] at (-.12,-2.945) {Q10};
\fill[metablue] (0.0000,-3.0600) rectangle (3.3250,-2.8300);
\node[text=white] at (1.6625,-2.9450) {4};
\fill[metablue!48] (3.3250,-3.0600) rectangle (6.6500,-2.8300);
\node[text=black] at (4.9875,-2.9450) {4};
\node[anchor=north] at (0.0,-3.30) {0\%};
\node[anchor=north] at (1.6625,-3.30) {25\%};
\node[anchor=north] at (3.325,-3.30) {50\%};
\node[anchor=north] at (4.9875,-3.30) {75\%};
\node[anchor=north] at (6.65,-3.30) {100\%};
\end{scope}
\begin{scope}[shift={(8.6,0)}]
\node[anchor=west,font=\small\bfseries] at (-.65,.55) {Teachers (n=6): global mean unrecoverable};
\node[anchor=east] at (-.12,0.115) {Q1};
\fill[metablue!48] (0.0000,0.0000) rectangle (1.1083,0.2300);
\node[text=black] at (0.5542,0.1150) {1};
\fill[black!15] (1.1083,0.0000) rectangle (3.3250,0.2300);
\node[text=black] at (2.2167,0.1150) {2};
\fill[metaorange!45] (3.3250,0.0000) rectangle (5.5417,0.2300);
\node[text=black] at (4.4333,0.1150) {2};
\draw[densely dotted,thick] (5.5417,0.0000) rectangle (6.65,0.2300);
\node at (6.0958,0.1150) {?};
\node[anchor=east] at (-.12,-0.22500000000000003) {Q2};
\fill[metablue] (0.0000,-0.3400) rectangle (2.2167,-0.1100);
\node[text=white] at (1.1083,-0.2250) {2};
\fill[metablue!48] (2.2167,-0.3400) rectangle (5.5417,-0.1100);
\node[text=black] at (3.8792,-0.2250) {3};
\fill[black!15] (5.5417,-0.3400) rectangle (6.6500,-0.1100);
\node[text=black] at (6.0958,-0.2250) {1};
\node[anchor=east] at (-.12,-0.5650000000000001) {Q3};
\fill[black!15] (0.0000,-0.6800) rectangle (1.1083,-0.4500);
\node[text=black] at (0.5542,-0.5650) {1};
\fill[metaorange!45] (1.1083,-0.6800) rectangle (5.5417,-0.4500);
\node[text=black] at (3.3250,-0.5650) {4};
\fill[metaorange] (5.5417,-0.6800) rectangle (6.6500,-0.4500);
\node[text=white] at (6.0958,-0.5650) {1};
\node[anchor=east] at (-.12,-0.905) {Q4};
\fill[metablue] (0.0000,-1.0200) rectangle (3.3250,-0.7900);
\node[text=white] at (1.6625,-0.9050) {3};
\fill[metablue!48] (3.3250,-1.0200) rectangle (5.5417,-0.7900);
\node[text=black] at (4.4333,-0.9050) {2};
\fill[metaorange] (5.5417,-1.0200) rectangle (6.6500,-0.7900);
\node[text=white] at (6.0958,-0.9050) {1};
\node[anchor=east] at (-.12,-1.245) {Q5};
\fill[black!15] (0.0000,-1.3600) rectangle (1.1083,-1.1300);
\node[text=black] at (0.5542,-1.2450) {1};
\fill[metaorange!45] (1.1083,-1.3600) rectangle (4.4333,-1.1300);
\node[text=black] at (2.7708,-1.2450) {3};
\fill[metaorange] (4.4333,-1.3600) rectangle (6.6500,-1.1300);
\node[text=white] at (5.5417,-1.2450) {2};
\node[anchor=east] at (-.12,-1.5850000000000002) {Q6};
\fill[metablue] (0.0000,-1.7000) rectangle (2.2167,-1.4700);
\node[text=white] at (1.1083,-1.5850) {2};
\fill[metablue!48] (2.2167,-1.7000) rectangle (5.5417,-1.4700);
\node[text=black] at (3.8792,-1.5850) {3};
\fill[black!15] (5.5417,-1.7000) rectangle (6.6500,-1.4700);
\node[text=black] at (6.0958,-1.5850) {1};
\node[anchor=east] at (-.12,-1.925) {Q7};
\fill[metaorange!45] (0.0000,-2.0400) rectangle (1.1083,-1.8100);
\node[text=black] at (0.5542,-1.9250) {1};
\fill[metaorange] (1.1083,-2.0400) rectangle (6.6500,-1.8100);
\node[text=white] at (3.8792,-1.9250) {5};
\node[anchor=east] at (-.12,-2.265) {Q8};
\fill[metablue] (0.0000,-2.3800) rectangle (3.3250,-2.1500);
\node[text=white] at (1.6625,-2.2650) {3};
\fill[metablue!48] (3.3250,-2.3800) rectangle (5.5417,-2.1500);
\node[text=black] at (4.4333,-2.2650) {2};
\fill[black!15] (5.5417,-2.3800) rectangle (6.6500,-2.1500);
\node[text=black] at (6.0958,-2.2650) {1};
\node[anchor=east] at (-.12,-2.605) {Q9};
\fill[metablue!48] (0.0000,-2.7200) rectangle (1.1083,-2.4900);
\node[text=black] at (0.5542,-2.6050) {1};
\fill[black!15] (1.1083,-2.7200) rectangle (2.2167,-2.4900);
\node[text=black] at (1.6625,-2.6050) {1};
\fill[metaorange!45] (2.2167,-2.7200) rectangle (5.5417,-2.4900);
\node[text=black] at (3.8792,-2.6050) {3};
\fill[metaorange] (5.5417,-2.7200) rectangle (6.6500,-2.4900);
\node[text=white] at (6.0958,-2.6050) {1};
\node[anchor=east] at (-.12,-2.945) {Q10};
\fill[metablue] (0.0000,-3.0600) rectangle (4.4333,-2.8300);
\node[text=white] at (2.2167,-2.9450) {4};
\fill[metablue!48] (4.4333,-3.0600) rectangle (6.6500,-2.8300);
\node[text=black] at (5.5417,-2.9450) {2};
\node[anchor=north] at (0.0,-3.30) {0\%};
\node[anchor=north] at (1.6625,-3.30) {25\%};
\node[anchor=north] at (3.325,-3.30) {50\%};
\node[anchor=north] at (4.9875,-3.30) {75\%};
\node[anchor=north] at (6.65,-3.30) {100\%};
\end{scope}
\fill[metablue] (0.0,-4.12) rectangle (0.22,-3.92);
\node[anchor=west] at (0.3,-4.02) {Strongly disagree};
\fill[metablue!48] (3.05,-4.12) rectangle (3.27,-3.92);
\node[anchor=west] at (3.3499999999999996,-4.02) {Disagree};
\fill[black!15] (6.1,-4.12) rectangle (6.319999999999999,-3.92);
\node[anchor=west] at (6.3999999999999995,-4.02) {Neutral};
\fill[metaorange!45] (9.149999999999999,-4.12) rectangle (9.37,-3.92);
\node[anchor=west] at (9.45,-4.02) {Agree};
\fill[metaorange] (12.2,-4.12) rectangle (12.42,-3.92);
\node[anchor=west] at (12.5,-4.02) {Strongly agree};
\end{tikzpicture}
\caption{Audited SUS item frequencies. Bar segments show counts; widths use each group\textquotesingle s expected denominator. The dotted teacher Q1 segment marks one undocumented response and is not imputed. Negative items (even numbers) are reverse scored when calculating SUS. Frequencies alone do not recover respondent-level variability.}
\label{fig:results}
\end{figure*}

The teacher chart contains five responses on item~1, although E4 included six teachers. The other nine items each contain six responses. An overall six-person mean cannot be recovered as a complete SUS result from this material. We do not assign a response to the missing position, substitute an item mean, or present 81.25 as confirmed. Figure~\ref{fig:results} and the supplement preserve the item distributions so that this limitation remains visible.

Item distributions still describe the response patterns without providing a defensible ranking of the two groups. For students, five of eight agreed or strongly agreed that the system was easy to use (item~3), and five expected that most people would learn it quickly (item~7). On the negatively worded item~8, seven disagreed or strongly disagreed that it was cumbersome; all eight disagreed or strongly disagreed that they needed to learn many things before using it (item~10). These favorable patterns coexist with more mixed responses about integration, consistency and confidence. For teachers, five of six agreed or strongly agreed with ease of use on item~3, and all six agreed or strongly agreed with rapid learnability on item~7. The incomplete first item prevents these item-level observations from becoming a verified global score.

SUS measures perceived usability, and its transformed values are not percentages of successful task performance \parencite{Lewis2018}. A response about learning to operate the system concerns interface learnability rather than learning software engineering. The student value is therefore reported descriptively, without using a single benchmark as an automatic acceptance decision. We calculate neither dispersion nor inferential comparisons: item marginals lack the respondent-level associations needed for those analyses, and the teachers' individual reproduction task differs from the students' paired interpretation task.

\subsection{Documented experiences and barriers}
\label{sec:e4-experiences}

The open-response summary shows that usability was uneven. The thesis classified ease-of-use feedback for one teacher and five students, and intuitive interaction for five students. It also recorded difficulty of use for four teachers and one student, a viewing difficulty for one teacher, and student comments categorized as chaotic interaction, difficulty interacting and a need for more adaptation time. Table~\ref{tab:e4-feedback} preserves these categories. A person could contribute to more than one category, and the available summary does not establish exhaustive coding rules. The counts should not be added to declare that a particular proportion of all participants was satisfied.

\begin{table}[t]
\centering
\caption{Documented E4 experience categories. T: teachers; S: students. A dash reproduces an empty source cell. Categories may overlap.}
\label{tab:e4-feedback}
\small
\begin{tabularx}{\columnwidth}{@{}Yrr@{}}
\toprule
Source category & T & S \\
\midrule
Attractive for students & 1 & -- \\
Ease of use & 1 & 5 \\
Good experience & 1 & 1 \\
Intuitive & -- & 5 \\
Quick to learn & -- & 1 \\
Chaotic & -- & 1 \\
Difficulty interacting & -- & 1 \\
Difficulty of use & 4 & 1 \\
Difficulty of viewing & 1 & -- \\
More time to adapt & -- & 1 \\
\bottomrule
\end{tabularx}
\end{table}

The requested changes identify where interaction competed with the educational task. Participants asked for a tutorial on interacting with and manipulating elements, alternative views such as a top view, a hierarchical view of classes, and positioning classes with the mouse rather than a slider. These requests concern orientation, visibility and spatial editing. The current synthesis interprets them as design priorities: a shared modelling space must help learners attend to class structure and relationships while keeping movement and object manipulation manageable. The requests alone do not show that implementing these features would improve learning or eliminate the reported difficulties.

The positive educational comments expressed perceived possibilities. Teacher P2 anticipated greater student interest and integration between teachers and students, and the ability to create worlds related to the subject. Teacher P6 described scenarios for understanding software design and architecture and using the third dimension in modelling. Student S6 anticipated that an engaging virtual world could draw attention to otherwise theoretical material. These extracts support an account of perceived utility and reasons for further investigation. They were not accompanied by a measure of attention, sustained engagement, course adoption or conceptual change.

The unfavorable comments are equally relevant to feasibility. Participants described complexity in creating classes, limitations in user experience, the time teachers might spend producing diagrams, potential student distraction and equipment limitations. A technically available feature may consequently remain costly to use within a lesson. In particular, the time required to build and arrange a model matters when modelling is meant to support discussion of software concepts. The possibility of distraction also qualifies the expectation that visual appeal automatically supports the intended learning activity.

\subsection{What the evaluations establish}

Together, the studies document a working route from educational design considerations to reusable modelling facilities and situated desktop use. E1--E3 identify perceived relevance alongside gaps in specificity, notation and instructional support. E4 adds experiences of entering a shared world and constructing class diagrams under remote conditions, with concrete feedback about navigation and manipulation. The pilot demonstrates a limited feedback-to-revision cycle; the main study supplies priorities for the next cycle.

The evidence supports formative claims about feasibility, perceived usability and barriers. It does not establish that participants learned more than with a conventional modelling tool, collaborated more effectively, remained engaged over a course, or adopted the platform after the session. No classroom learning outcome was measured in E4, and the antecedent pre/post study demonstrated no gain. The incomplete teacher aggregates and conflicting student SUS summaries further limit numerical conclusions. These boundaries determine the design implications developed next and the requirements for a stronger subsequent educational evaluation.

\section{Discussion and limitations}
\label{sec:discussion}

\subsection{What the educational design contributes}
For RQ1, MetaSEE provides a documented route from educational and technological requirements to an instantiation organized around UML modeling. Its value lies in making the design decisions inspectable: what learners are meant to do, what representation they manipulate, how they encounter other users, and which interaction demands the environment introduces. The framework and feature model articulate choices, while VirtualUML and the platform make a narrower subset operational. Treating these artifacts as interchangeable would obscure their different evidential status. A guideline judged relevant is not an implemented capability, and a capability visible in a prototype is not a demonstrated educational benefit.

The affordance perspective of \textcite{DalgarnoLee2010} helps interpret this connection. A shared spatial representation can provide a common object for discussion, and an environment can allow learners to explore perspectives or rehearse activity. In MetaSEE, the connection is clearest where pairs work on a class model that both can inspect. The educational mechanism remains conditional: students need to explain and negotiate modeling choices, rather than merely move objects. The present evidence establishes the availability of the activity and records participants' perceptions. It does not establish whether explanation quality, conceptual understanding, or transfer improved.

This distinction sharpens the relationship to neighboring work. SENEM addresses requirements and iterative evaluation for educational metaverse development \parencite{Pentangelo2024}, whereas MetaSEE contributes a traceable case linking design artifacts to a particular modeling activity and its barriers. Collaborative UML work such as \textcite{Yigitbas2023} makes a feature-level comparison more informative than a generic claim that using 3D is novel. The useful questions concern representations, editing actions, collaboration conditions, and evaluation measures. The comparison in the related-work section identifies those differences without ranking platforms that were not evaluated under common conditions. Similarly, the recent computing-education review \parencite{Janani2026} provides context for assessing applications and evidence, not independent validation of MetaSEE's outcomes.

\subsection{How formative findings should change the design}
For RQ2, E1--E3 identify refinements at three distinct levels. The framework needs clearer separation of educational goals from development decisions and more concrete guidance for applying its prompts. The feature model needs explanations connecting device capabilities to configuration choices and a clearer account of what its coverage excludes. VirtualUML needs support for spatial coordinates and model complexity, alongside fuller notation where instruction requires it. These implications preserve the participants' criticism rather than recasting all feedback as approval. They also avoid assuming that more features are always the next educational priority: a smaller representation that students can interpret may be more useful than a broader notation they cannot reliably manipulate.

For RQ3, the platform evidence points toward navigation and readability as conditions of participation. An introductory sequence should let users move, select a class, edit its content, and recover their view before they undertake a modeling problem. Alternative or top views could help users inspect relationships when spatial arrangements become hard to read. Those suggestions require implementation and evaluation; they are not features demonstrated by this study. The teacher requests and the student responses also caution against using rapid familiarization or enthusiasm as substitutes for observing task success. Familiarity with games may be relevant to interaction, but these data cannot establish its causal effect.

\begin{table*}[t]
\caption{Design implications derived from documented findings. Proposed changes and future measures are recommendations, not implemented or observed outcomes.}
\label{tab:implications}
\small
\begin{tabularx}{\textwidth}{@{}P{2.1cm}YYY@{}}
\toprule
Evidence source & Documented issue & Design implication & Evidence needed in a subsequent study\\
\midrule
E1 framework & Generic prompts; divided views on clarity & Separate learning goals, learner context, and implementation choices; add worked examples & Quality and completeness of requirements produced with and without guidance\\
E2 feature model & Completeness unresolved; device detail requested & Explain capability constraints and configuration boundaries & Correctness of generated configurations and effort under defined development tasks\\
E3 component & Coordinate guidance and richer notation requested & Support placement and legibility; specify supported UML subset & Task errors and model readability as diagram complexity increases\\
E4 platform & Navigation, interaction, and viewing difficulties & Add guided practice, clearer selection feedback, and alternative views & Unassisted completion, recovery from errors, and time spent on modeling versus navigation\\
E4 educational expectations & Potential collaboration and distraction & Provide explicit pair roles and prompts for explaining decisions & Recorded modeling dialogue, contribution patterns, model quality, and delayed understanding\\
\bottomrule
\end{tabularx}
\end{table*}

These priorities suggest a staged evaluation strategy. First, establish whether novice users can complete a defined model without excessive interaction support. Second, examine how pairs distribute work and justify their class and relationship decisions. Third, compare educational outcomes under equivalent instructional content, duration, and assessment. That sequence would distinguish an interface improvement from a pedagogical effect. The original GQM orientation can support the definition of these measures, but the proposed studies are future work rather than an extension of the present dataset.

\subsection{Interpretation and validity boundaries}
The numerical audit changes what can be concluded about usability. The student score of 67.50 is conditional on the documented marginal counts; it does not recover the original questionnaire file. The teacher distribution is incomplete. Neither historical mean can therefore be used to certify satisfactory usability or teacher superiority. SUS is a measure of perceived usability on a transformed scale, not a percentage of correct actions or a learning assessment \parencite{Brooke1996,Lewis2018}. Without within-person responses, reliability, dispersion, confidence intervals, and associations with profile variables cannot be recovered. A transparent account of the discrepancy is more informative than selecting whichever reported number is more favorable.

The samples are small, selected through convenience or voluntary recruitment, and tied to a limited institutional and technological context. They can reveal problems that deserve design attention, but do not estimate the prevalence of those problems in a broader student or teacher population. The same 12 developers participated in E1 and E2, and other cross-study overlap cannot be resolved. Summing evaluation counts would overstate the number of independent people. Different instruments and exposure conditions also prevent treating the four evaluations as replications or pooling their favorable responses.

Within E4, teachers reproduced a supplied class diagram individually, while students modeled a written scenario in pairs. Task demands, social support, and participant profiles therefore vary together. Remote conferencing and communication tools were part of the procedure, so successful communication cannot be attributed exclusively to the platform. Short exploration and single-session participation also leave novelty, sustained use, and the need for support unresolved. The findings apply to the documented desktop interaction. They do not establish usability with headsets, controllers, gestures, or other immersive modalities.

The qualitative synthesis is constrained by the surviving record. It uses the thesis's reported categories and extracts, without the complete response corpus or recordings needed for an independent coding process. Categories may overlap; blank cells do not establish verified zero responses. Frequency of a reported category does not demonstrate causal importance or thematic saturation. Researcher involvement in building and evaluating the prototype may also shape both the questions and interpretation. These limitations make unfavorable observations particularly useful for checking an overly positive account, while not turning individual criticisms into population estimates.

Finally, the current pedagogical mapping and research questions organize an existing record retrospectively. They were not a preregistered test of an affordance theory. The preliminary requirements-learning activity is separate from E4 and showed no increase in its four participants' test results; it cannot supply the missing evidence of effectiveness. Learning analytics and interoperability among independent applications remain proposed extensions. Administrative verification of ethics approval, consent records, and publication overlap is also outstanding in the submission package. Together, these boundaries support a contribution in educational design and formative evaluation while specifying what a subsequent classroom study must establish.

\section{Conclusion}
\label{sec:conclusion}
MetaSEE illustrates how educational requirements can be connected to a feature model, a reusable UML representation, and a browser-based collaborative environment for software engineering education. The article's contribution is the traceability of those design choices to documented formative evidence. It makes the implemented prototype, the conceptual design artifacts, and future extensions explicit so that each can be assessed on appropriate grounds.

The complementary evaluations identify both perceived usefulness and unresolved design concerns. The framework needs more concrete and clearly organized guidance; the feature model needs clearer capability boundaries; and the UML component needs better support for spatial interaction and richer notation. In the main platform evaluation, six teachers worked individually and eight students worked in four pairs. Their feedback indicates potential value in shared modeling alongside navigation, manipulation, visualization, and distraction barriers. These observations concern a desktop or notebook experience using keyboard and mouse.

The evidence audit qualifies the quantitative conclusion. The student SUS mean derivable from documented item frequencies is 67.50, in conflict with the historical 79.38. The teacher chart lacks one response in Q1, preventing recovery of a complete group mean. These discrepancies are retained transparently, with reproducible aggregate calculations in the accompanying materials. They preclude using the previous means as confirmation of favorable platform usability.

Further development should prioritize guided entry into the modeling task, clear selection and editing feedback, readable model views, and explicit support for collaborative explanation. Subsequent evaluations should preserve individual response data and examine unassisted interaction, model quality, and the reasoning expressed by pairs before testing learning outcomes under comparable instructional conditions. Longer classroom use is needed to assess sustained participation and practical teaching integration. Studies using immersive devices and independent applications would address different capabilities and require separate evidence.

The present record supports an account of design feasibility, participant perceptions, and actionable barriers. It leaves learning gains, sustained engagement, adoption, and broad interoperability as questions for future empirical work.

\section*{Declarations}
\textbf{Ethics and consent.} The study documentation includes a consent instrument describing voluntary adult participation, confidentiality, withdrawal, and aggregate reporting; the E4 protocol required consent before participation. Ethics approval has been reported by the authors, but the committee, approval identifier, date, and coverage of E1--E4 remain to be checked against the approval document before submission. Signed individual consent records were not available for this documentary synthesis.

\textbf{Data availability.} This article uses the documented evaluation protocols, aggregate charts, tables, and participant statements in the source research record. The individual questionnaire responses, interview transcripts, and recordings were unavailable to the present synthesis. The accompanying supplement and audit files provide the extracted aggregates, calculations, provenance, and unresolved discrepancies. They cannot support participant-level statistical reanalysis.

\textbf{Funding and competing interests.} Funding sources, grant identifiers, and competing-interest declarations require confirmation by all authors before submission. No absence of funding or conflicts is inferred from silence in the source record.

\textbf{Author contributions.} The author-specific CRediT statement accompanies the separate title-page draft and requires approval by all authors.

\textbf{Declaration of generative AI and AI-assisted technologies in the manuscript preparation process.} OpenAI Codex was used to assist with documentary extraction, literature searching, English drafting, LaTeX preparation, aggregate calculation code, and consistency checking. The authors must review and edit the resulting text and take responsibility for the final submitted content. No new participant observations or responses were generated.

\balance
\printbibliography

@article{DalgarnoLee2010,
 author={Dalgarno, Barney and Lee, Mark J. W.},
 title={What are the learning affordances of {3-D} virtual environments?},
 journaltitle={British Journal of Educational Technology},
 date={2010}, volume={41}, number={1}, pages={10--32},
 doi={10.1111/j.1467-8535.2009.01038.x}
}

@article{Fowler2015,
 author={Fowler, Chris},
 title={Virtual reality and learning: Where is the pedagogy?},
 journaltitle={British Journal of Educational Technology},
 date={2015}, volume={46}, number={2}, pages={412--422},
 doi={10.1111/bjet.12135}
}

@article{Janani2026,
 author={Janani, Azin S. and Yang, Yanzhuo and Lu, Xundi and Liang, Hai-Ning and Baghaei, Nilufar},
 title={Virtual reality applications in computer science education: Scoping review},
 journaltitle={Computers \& Education: X Reality},
 date={2026}, volume={8}, eid={100159},
 doi={10.1016/j.cexr.2026.100159}
}

@article{Pentangelo2024,
 author={Pentangelo, Viviana and Di Dario, Dario and Lambiase, Stefano and Ferrucci, Filomena and Gravino, Carmine and Palomba, Fabio},
 title={{SENEM}: A software engineering-enabled educational metaverse},
 journaltitle={Information and Software Technology},
 date={2024}, volume={174}, eid={107512},
 doi={10.1016/j.infsof.2024.107512}
}

@article{Jovanovic2022,
 author={Jovanović, Aleksandar and Milosavljević, Aleksandar},
 title={{VoRtex} metaverse platform for gamified collaborative learning},
 journaltitle={Electronics}, date={2022}, volume={11}, number={3}, eid={317},
 doi={10.3390/electronics11030317}
}

@article{Roy2023,
 author={Roy, Rita and Babakerkhell, Mohammad Dawood and Mukherjee, Subhodeep and Pal, Debajyoti and Funilkul, Suree},
 title={Development of a framework for metaverse in education: A systematic literature review approach},
 journaltitle={IEEE Access}, date={2023}, volume={11}, pages={57717--57734},
 doi={10.1109/ACCESS.2023.3283273}
}

@article{Yigitbas2023,
 author={Yigitbas, Enes and Gorissen, Simon and Weidmann, Nils and Engels, Gregor},
 title={Design and evaluation of a collaborative {UML} modeling environment in virtual reality},
 journaltitle={Software and Systems Modeling}, date={2023}, volume={22}, pages={1397--1425},
 doi={10.1007/s10270-022-01065-2}
}

@article{Mayor2021,
 author={Mayor, Jesus and López-Fernández, Daniel},
 title={{Scrum VR}: Virtual reality serious video game to learn {Scrum}},
 journaltitle={Applied Sciences}, date={2021}, volume={11}, number={19}, eid={9015},
 doi={10.3390/app11199015}
}

@article{LopezFernandez2023,
 author={López-Fernández, Daniel and Mayor, Jesus and Pérez, Jennifer and Gordillo, Aldo},
 title={Learning and motivational impact of using a virtual reality serious video game to learn {Scrum}},
 journaltitle={IEEE Transactions on Games}, date={2023}, volume={15}, number={3}, pages={430--439},
 doi={10.1109/TG.2022.3213127}
}

@article{BezzinaZammit2026,
 author={Bezzina, Andrea and Zammit, Joseph Paul},
 title={A comprehensive framework for designing metaverse-based learning for engineering education},
 journaltitle={Multimodal Technologies and Interaction}, date={2026}, volume={10}, number={7}, eid={74},
 doi={10.3390/mti10070074}
}

@inproceedings{Rodrigues2016,
 author={Rodrigues, Claudia Susie C. and Werner, Cláudia M. L. and Landau, Luiz},
 title={{VisAr3D}: An innovative {3D} visualization of {UML} models},
 booktitle={Proceedings of the 38th International Conference on Software Engineering Companion},
 date={2016}, pages={451--460}, publisher={Association for Computing Machinery},
 doi={10.1145/2889160.2889199}
}

@article{FernandesWerner2022,
 author={Fernandes, Filipe A. and Werner, Cláudia M. L.},
 title={A scoping review of the metaverse for software engineering education: Overview, challenges, and opportunities},
 journaltitle={PRESENCE: Virtual and Augmented Reality}, date={2022}, volume={31}, pages={107--146},
 doi={10.1162/pres_a_00371}
}

@article{FernandesFrameworks2023,
 author={Fernandes, Filipe Arantes and Rodrigues, Claudia Susie Camargo and Teixeira, Eldânae Nogueira and Werner, Cláudia M. L.},
 title={Immersive learning frameworks: A systematic literature review},
 journaltitle={IEEE Transactions on Learning Technologies}, date={2023}, volume={16}, number={5}, pages={736--747},
 doi={10.1109/TLT.2023.3242553}
}

@article{Garousi2019,
 author={Garousi, Vahid and Giray, Gorkem and Tuzun, Eray},
 title={Understanding the knowledge gaps of software engineers: An empirical analysis based on {SWEBOK}},
 journaltitle={ACM Transactions on Computing Education}, date={2019-11}, volume={20}, number={1}, eid={3},
 doi={10.1145/3360497}
}

@article{Teyseyre2009,
 author={Teyseyre, Alfredo R. and Campo, Marcelo R.},
 title={An overview of {3D} software visualization},
 journaltitle={IEEE Transactions on Visualization and Computer Graphics},
 date={2009}, volume={15}, number={1}, pages={87--105},
 doi={10.1109/TVCG.2008.86}
}

@article{Lewis2018,
 author={Lewis, James R.}, title={The {System Usability Scale}: Past, present, and future},
 journaltitle={International Journal of Human--Computer Interaction},
 date={2018}, volume={34}, number={7}, pages={577--590},
 doi={10.1080/10447318.2018.1455307}
}

@incollection{Brooke1996,
 author={Brooke, John}, title={{SUS}: A ``quick and dirty'' usability scale},
 editor={Jordan, Patrick W. and Thomas, Bruce and Weerdmeester, Bernard A. and McClelland, Ian L.},
 booktitle={Usability Evaluation in Industry}, date={1996}, pages={189--194},
 publisher={Taylor \& Francis}, doi={10.1201/9781498710411-35}
}

@article{Hevner2004,
 author={Hevner, Alan R. and March, Salvatore T. and Park, Jinsoo and Ram, Sudha},
 title={Design science in information systems research},
 journaltitle={MIS Quarterly}, date={2004}, volume={28}, number={1}, pages={75--105},
 doi={10.2307/25148625}
}

@article{Peffers2007,
 author={Peffers, Ken and Tuunanen, Tuure and Rothenberger, Marcus A. and Chatterjee, Samir},
 title={A design science research methodology for information systems research},
 journaltitle={Journal of Management Information Systems},
 date={2007}, volume={24}, number={3}, pages={45--77},
 doi={10.2753/MIS0742-1222240302}
}

@incollection{Basili1994,
 author={Basili, Victor R. and Caldiera, Gianluigi and Rombach, H. Dieter},
 title={The goal question metric approach},
 editor={Marciniak, John J.}, booktitle={Encyclopedia of Software Engineering},
 date={1994}, volume={1}, pages={528--532}, publisher={John Wiley \& Sons},
 url={https://www.cs.umd.edu/~basili/papers.html}
}

@article{Davis1989,
 author={Davis, Fred D.},
 title={Perceived usefulness, perceived ease of use, and user acceptance of information technology},
 journaltitle={MIS Quarterly}, date={1989}, volume={13}, number={3}, pages={319--340},
 doi={10.2307/249008}
}

@phdthesis{FernandesThesis2023,
 author={Fernandes, Filipe Arantes},
 title={{MetaSEE}: An Approach to Enable the Metaverse-based Software Engineering Education},
 school={Federal University of Rio de Janeiro},
 date={2023-05}, langid={english}
}
\end{document}